\documentclass[prb,twocolumn,showpacs,preprintnumbers,superscriptaddress,amssymb]{revtex4}

\usepackage[fleqn]{amsmath}
\usepackage{graphicx}
\usepackage{dcolumn}
\usepackage{bm}
\usepackage{epstopdf}
\usepackage{xcolor}
\usepackage[normalem]{ulem}

\begin{document}


\title{Bilayer-induced asymmetric quantum Hall effect in epitaxial graphene}

\author{Andrea~Iagallo}
\affiliation{NEST, Istituto Nanoscienze--CNR and Scuola Normale Superiore, \\Piazza San Silvestro 12, 56127 Pisa,  Italy}

\author{Shinichi~Tanabe}
\affiliation{NTT Basic Research Laboratories, NTT Corporation, 3-1 Morinosato Wakamiya, Atsugi, Kanagawa, 243-0198, Japan}

\author{Stefano~Roddaro}
\affiliation{NEST, Istituto Nanoscienze--CNR and Scuola Normale Superiore, \\Piazza San Silvestro 12, 56127 Pisa,  Italy}

\author{Makoto~Takamura}
\affiliation{NTT Basic Research Laboratories, NTT Corporation, 3-1 Morinosato Wakamiya, Atsugi, Kanagawa, 243-0198, Japan}

\author{Yoshiaki~Sekine}
\affiliation{NTT Basic Research Laboratories, NTT Corporation, 3-1 Morinosato Wakamiya, Atsugi, Kanagawa, 243-0198, Japan}

\author{Hiroki~Hibino}
\affiliation{NTT Basic Research Laboratories, NTT Corporation, 3-1 Morinosato Wakamiya, Atsugi, Kanagawa, 243-0198, Japan}

\author{Vaidotas Miseikis}
\affiliation{Center for Nanotechnology Innovation @ NEST, Istituto Italiano di Tecnologia, Piazza San Silvestro 12, 56127 Pisa, Italy}

\author{Camilla Coletti}
\affiliation{Center for Nanotechnology Innovation @ NEST, Istituto Italiano di Tecnologia, Piazza San Silvestro 12, 56127 Pisa, Italy}

\author{Vincenzo Piazza}
\affiliation{Center for Nanotechnology Innovation @ NEST, Istituto Italiano di Tecnologia, Piazza San Silvestro 12, 56127 Pisa, Italy}

\author{Fabio Beltram}
\affiliation{NEST, Istituto Nanoscienze--CNR and Scuola Normale Superiore, \\Piazza San Silvestro 12, 56127 Pisa,  Italy}
\affiliation{Center for Nanotechnology Innovation @ NEST, Istituto Italiano di Tecnologia, Piazza San Silvestro 12, 56127 Pisa, Italy}

\author{Stefan~Heun}
\email{stefan.heun@nano.cnr.it}
\affiliation{NEST, Istituto Nanoscienze--CNR and Scuola Normale Superiore, \\Piazza San Silvestro 12, 56127 Pisa,  Italy}



\date{\today}

\begin{abstract}

The transport properties of epitaxial graphene on SiC(0001) at quantizing magnetic fields are investigated. Devices patterned perpendicularly to SiC terraces clearly exhibit bilayer inclusions distributed along the substrate step edges. We show that the transport properties in the quantum Hall regime are heavily affected by the presence of bilayer inclusions, and observe a significant departure from the conventional quantum Hall characteristics. A quantitative model involving enhanced inter-channel scattering mediated by the presence of bilayer inclusions is presented that successfully explains the observed symmetry properties.

\end{abstract}

\maketitle

\section{\label{secIntro} Introduction}

Graphene has rapidly become an exciting material in fundamental research and the high expectations on its device applications are currently stimulating a large effort towards its large-scale production in order to meet potential market needs. Passing from $\mu$m--sized high--mobility exfoliated graphene,\cite{NovoselovScience2004,BolotinNat2009} an approach which is suitable for laboratory research, to wafer size while preserving the desired electronic properties is not a trivial matter. In the development of a scalable graphene-based electronics, the quality of exfoliated graphene is used as a benchmark that sets the standard for the properties to be made available to the electronics industry. Two main routes are pursued in the quest for monolithic integration of graphene: epitaxial growth of graphene by chemical vapor deposition (CVD) on catalytic metals\cite{WeiNL2009,KinNat2009,CaoNL2010} and thermal decomposition of silicon carbide (SiC).\cite{EmtsevNatMater2009,BergerScience2006} Both techniques recently achieved at wafer scale mobility and uniformity levels comparable to those of exfoliated graphene.\cite{ShenJAP2012,TanabePRB2011} As a major drawback, graphene produced by CVD must be transferred onto an insulating substrate for its use in electronics applications. On the contrary, the SiC substrate is insulating, and this in principle offers a clear technological advantage over other methods. Using this technique, graphene field-effect transistors grown on a two-inch SiC wafer and exhibiting a maximum frequency up to 100 GHz were demonstrated.\cite{LinScience2010}

In the SiC-based process, graphene layers are obtained by annealing SiC wafers at high temperature (above $1200\,^{\circ}\mathrm{C}$).\cite{VarchonPRL2007} This induces a decomposition of the SiC, the sublimation of Si atoms, and the formation of a graphene monolayer. The presence of a slight wafer miss-cut causes a number of atomically sharp step edges separated by flat terraces across the crystal surface \cite{SeyllerSurSci2006} where nucleation of multilayer graphene domains is favored.\cite{TanakaPRB2010}

As a result, typical large-area graphene devices on SiC contain epitaxial monolayer graphene on top of the vicinal surface of the SiC substrate, with narrow multilayer inclusions which run along the SiC step edges. In devices grown on the Si--face of SiC (SiC(0001)), a high control can be achieved on the number of graphene layers, and thus multilayer regions are mainly composed of bilayer graphene. These bilayer areas are a source of electron scattering and inhomogeneity in carrier density, and recently attracted attention since they could hinder large-scale integration.

From the experimental side, the relation between magnetotransport properties and orientation of SiC step edges is not clear. In early work, the electrical resistance of graphene on SiC was found to be larger when measured in the direction perpendicular to the surface terraces compared to the parallel one.\cite{YakesNL2010} Analogously, the mobility of large--area Hall--bars perpendicular to the step edges was found to be significantly lower than in devices aligned parallel to them.\cite{TanabePRB2011} On the other hand, other investigations of epitaxial devices on SiC(0001) deliberately fabricated either on single terraces or crossing several step edges \cite{JobstPRB2010,JobstSSC2011} found no significant effect on the mobility and carrier density.

At larger magnetic fields, in devices in which the bilayer patches did not cross even the narrowest part of the Hall--bar the expected half--integer quantum Hall (QH) effect was observed irrespective of the orientation between the device and the substrate,\cite{JobstSSC2011,PallecchiSR2014} thus demonstrating that the monolayer graphene is continuous over the step edges. Anomalous QH traces were observed, however, in the presence of extended bilayer inclusions, which appear as continuous stripes intersecting the path of the propagating transport channels. Recently, a few experimental works focused on the influence of bilayer inclusions on the QH effect \cite{SchumannPRB2012,LofwanderPRB2013,YagerNL2013} and confirmed that bilayer domains crossing the Hall--bars induce shunting of edge channels and destroy the half--integer QH quantization expected for monolayer graphene. Additional transverse transport channels in the bilayer domains were put forward as possible mechanism causing shunting of the edge states, but a quantitative comparison with theory is still lacking.

In the present work, we investigate the influence of bilayer inclusions on the non-local transport in a Hall--bar in the QH regime. Our data show anomalous values of the quantized resistance and a peculiar asymmetric dependence on magnetic field which was not observed before, and that we fully account for in the framework of the B\"uttiker-Landauer model. Our data show the coexistence of the QH effect with different filling factors in both monolayer and bilayer grown on the same substrate.

\section{\label{secExpDetails} Experimental Details}

\begin{figure}[tbp]
\includegraphics[width=\columnwidth]{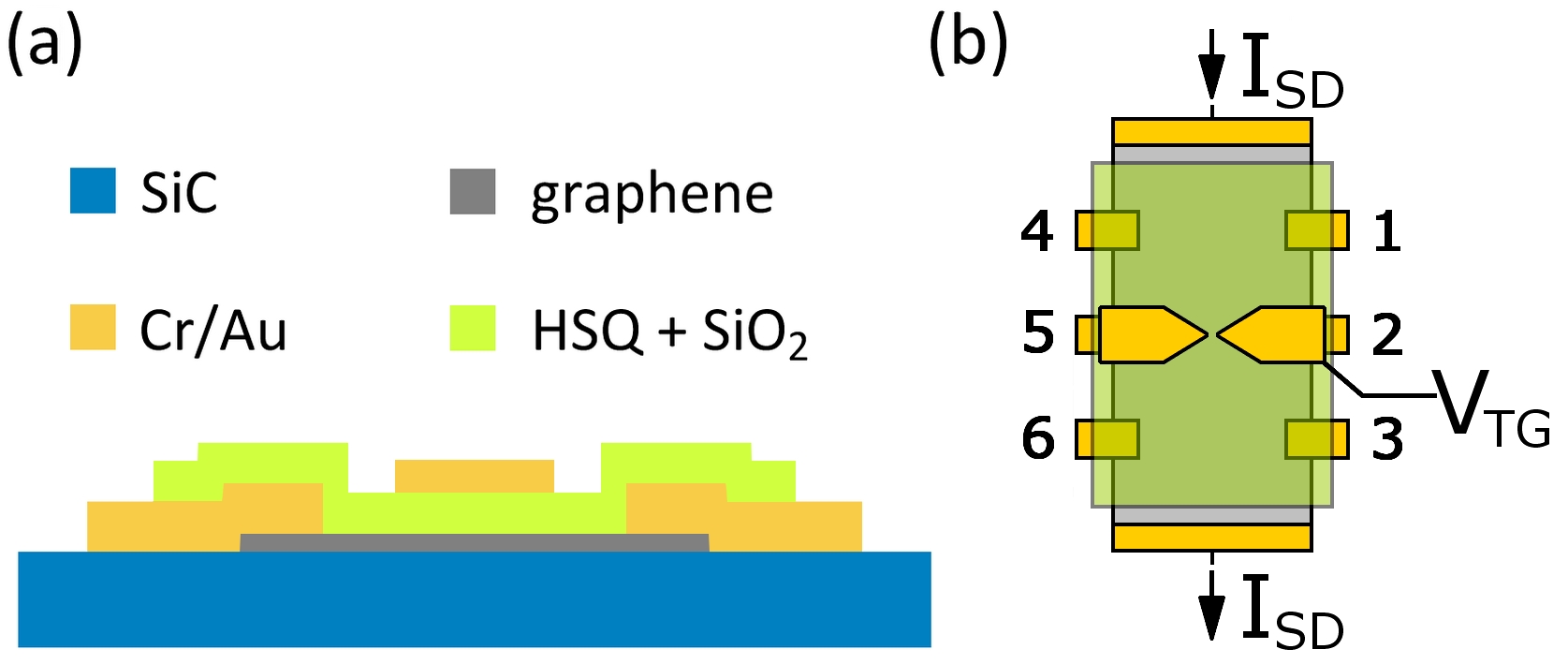}
\caption{(Color online) (a) Cross-section and (b) layout of the device. The bilayer dielectric comprising Hydrogen Silsequioxane (HSQ) and SiO$_{2}$ is shown as a single layer (green). The longitudinal and transversal resistances are acquired in a 4-point configuration at constant source-drain current $I_{SD}$.}
\label{fig1}
\end{figure}

The device investigated in this work is a Hall--bar (length~$\times$~width~$=$~300~$\mu$m~$\times$~50~$\mu$m) fabricated by standard optical lithography from an epitaxial graphene layer grown on a 6H-SiC(0001) substrate. The epitaxial graphene was grown by annealing 6H-SiC(0001) in Ar atmosphere at 100 Torr. The annealing temperature was approximately $1820\,^{\circ}\mathrm{C}$. A cross-section and a schematic of our device is displayed in Figs.~\ref{fig1}(a) and (b), respectively. The orientation of the Hall--bar was deliberately chosen to be perpendicular to the SiC step edges, in order to have the terraces running across the device from one side to the other. Ohmic contacts were made by a Cr/Au (5/250~nm) metallic layer. As can be seen in Fig.~\ref{fig1}(b), the ohmic contacts were not recessed as commonly done, but jutted out into the Hall--bar. A bilayer stack (140~nm of Hydrogen Silsequioxane (HSQ) and 40~nm of SiO$_{2}$)\cite{TanabeAPEX2010} was used as a dielectric. On top of this dielectric, Cr/Au (10/30~nm) electrodes in split-gate geometry were defined by e-beam lithography (not used in the experiments discussed here). We obtained similar results from another, nominally identical device.

Transport measurements were performed in a Heliox He$^{3}$ cryostat with a base temperature of 250~mK. Longitudinal and transverse resistances $R_{xx}=V_{xx}/I_{SD}$ ($xx=4-6,1-3$) and $R_{xy}=V_{xy}/I_{SD}$ ($xy=4-1,6-3$) were measured by standard lock-in technique in a 4-point configuration using small bias currents $I_{SD}\sim 10$~nA to avoid heating. Magnetic fields in the $0-9$~T range were used to characterize the device in the QH regime. The measured values of the carrier density and mobility are $n=3.4 \times 10^{11}$~cm$^{-2}$ and $\mu=4660$~cm$^{2}$/Vs, respectively.

Raman spectra were collected at room temperature, using a Renishaw Micro-Raman spectrometer employing a 532 nm laser excitation and a typical spot diameter of $<1\;\mu$m.

\section{\label{secExpRes} Results and Discussion}

\subsection{\label{secRaman} Micro-Raman layer topography}

Raman spectroscopy is a powerful tool to differentiate between mono-- and multilayer graphene.\cite{FerrariPRL2006} In the case of SiC substrates most studies concentrate on the so-called 2D peak at around 2700 cm$^{-1}$ and due to a double resonant scattering  process involving two phonons. It is the signal of choice since its frequency lies far away from the excitation structures of the underlying SiC substrate that make the detection of other graphene signatures more difficult.

The intensity, width and frequency of the 2D peak are sensitive to the number of layers.\cite{FerrariPRL2006} However, the 2D peak shows also a great sensitivity to strain and charge inhomogeneity: this make the assessment of layer number not conclusive. The ultimate signature of monolayer graphene is considered to be the line-shape of the 2D peak.\cite{GrafNL2007,RohrAPL2008l} In particular, there is general agreement that the peak of monolayer graphene can be fitted with a single Lorentzian, while a decomposition into four Lorentzians is necessary for bilayer graphene.\cite{LeeNL2008}

\begin{figure*}[t]
\includegraphics[width=\textwidth]{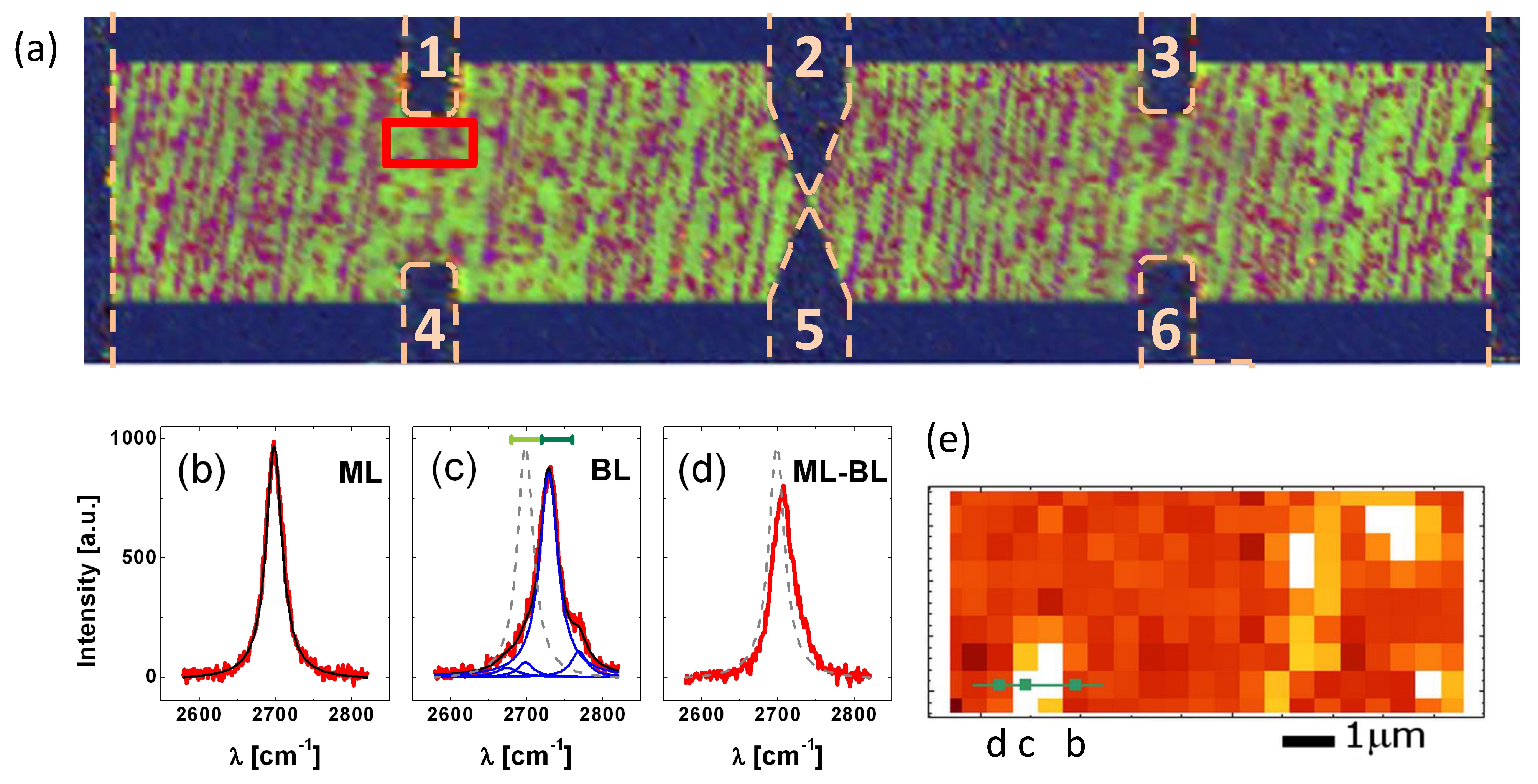}
\caption{(Color online) Raman topography of the device. (a) Composite map of Raman intensity integrated over the intervals $2680-2720$ cm$^{-1}$ (range $I$) and $2720-2760$ cm$^{-1}$ (range $II$). Large stripes of monolayer graphene (light green) lie on the SiC terraces, and are partially intersected by narrower bilayer domains (violet). An overlay indicating the ohmic contacts and the split-gates is also shown. (b)-(d) 2D peak extracted from the Raman spectra acquired at three different points of the device. Their positions are indicated in (e), which is a low-noise Raman intensity map, integrated over the $2720-2760$ cm$^{-1}$ range (bilayer), of a portion of the device marked by the red rectangle in (a). Monolayer domains (b) can be fitted with a single Lorentzian, while bilayer domains (c) require a four-Lorentzian fit. At the same time, the 2D peak of bilayer domains is red-shifted, which allows to identify the number of layer easily from the integrated intensity. The dashed lines in (c) and (d) repeat the monolayer curve from (b), to visualize the red shift, while the green bars in (c) indicate range $I$ (light green) and range $II$ (dark green). The RGB composite map in (a) contains range $I$ in the G channel, range $II$ in the B channel, and the sum of range $I$ and $II$ in the R channel.}
\label{fig2}
\end{figure*}

As a first step, we produced a fine Raman-scattering map to determine the layer topography of the entire device. Figure~\ref{fig2}(a) shows our result for a step size of 0.5~$\mu$m with integration times as long as 10 s for low-noise spectrum detection. As discussed in the following, for our device we were able to detect the number of graphene layers directly by integrating the scattering intensity in suitable ranges. A systematic analysis of the spectra acquired at several points across the device gives us confidence that we can identify two integration ranges corresponding to monolayer and bilayer graphene.

In order to obtain a Raman map, we first acquired the whole spectrum ($1300-2800\;cm^{-1}$) at each point and corrected it for the SiC contribution by subtracting a reference spectrum of the bare substrate. The resulting spectral data were then integrated over the ranges indicated in Figs.~\ref{fig2}(b-d), where three typical spectra acquired at different positions are shown, each one exhibiting different and peculiar characteristics. At point (b), the shape of the 2D peak is perfectly fitted by a single Lorentzian, thus demonstrating that monolayer graphene is present at this position. At point (c) the peak has a complex shape, and it is significantly red-shifted. Most importantly, the peak can be satisfactorily fitted only by using four Lorentzians, a clear signature of bilayer graphene. Point (d) shows the intermediate situation, where the laser spot covers an area where mostly monolayer graphene is present, but the bilayer contribution cannot be neglected. By exploiting the red-shift occurring for bilayer graphene, we can identify two separate integration ranges $2680-2720$ cm$^{-1}$ (range $I$) and $2720-2760$ cm$^{-1}$ (range $II$) for the detection of the number of graphene layers. We carefully checked that red-shifted peaks could be fitted by four Lorentzians only by analyzing the spectra at several locations, and we could assign range $I$ and range $II$ to monolayer and bilayer graphene, respectively. As an example, the points where the spectra in Figs.~\ref{fig2}(b-d) were acquired are shown in Fig.~\ref{fig2}(e), which is a low noise map of integrated intensity in range $II$, where white is high intensity and dark red is low intensity.

The Raman topography of the whole device was obtained first by calculating the two integrated intensity maps in the range $I$ and $II$, and then by combining them into a composite image using RGB channels. The final map is shown in Fig.~\ref{fig2}(a). In the figure, light green (violet) indicates a large integral value in range $I$ (range $II$) and thus the presence of monolayer (bilayer) graphene. The Hall--bar is intersected by tens of SiC step edges along which bilayer inclusions are present. Some of these inclusions are isolated islands, whereas other form long continuous stripes. In particular, some bilayer stripes connect one side of the device to the other.

\subsection{\label{secQH} Quantum Hall regime}

\begin{figure}[tbp]
\includegraphics[width=\columnwidth]{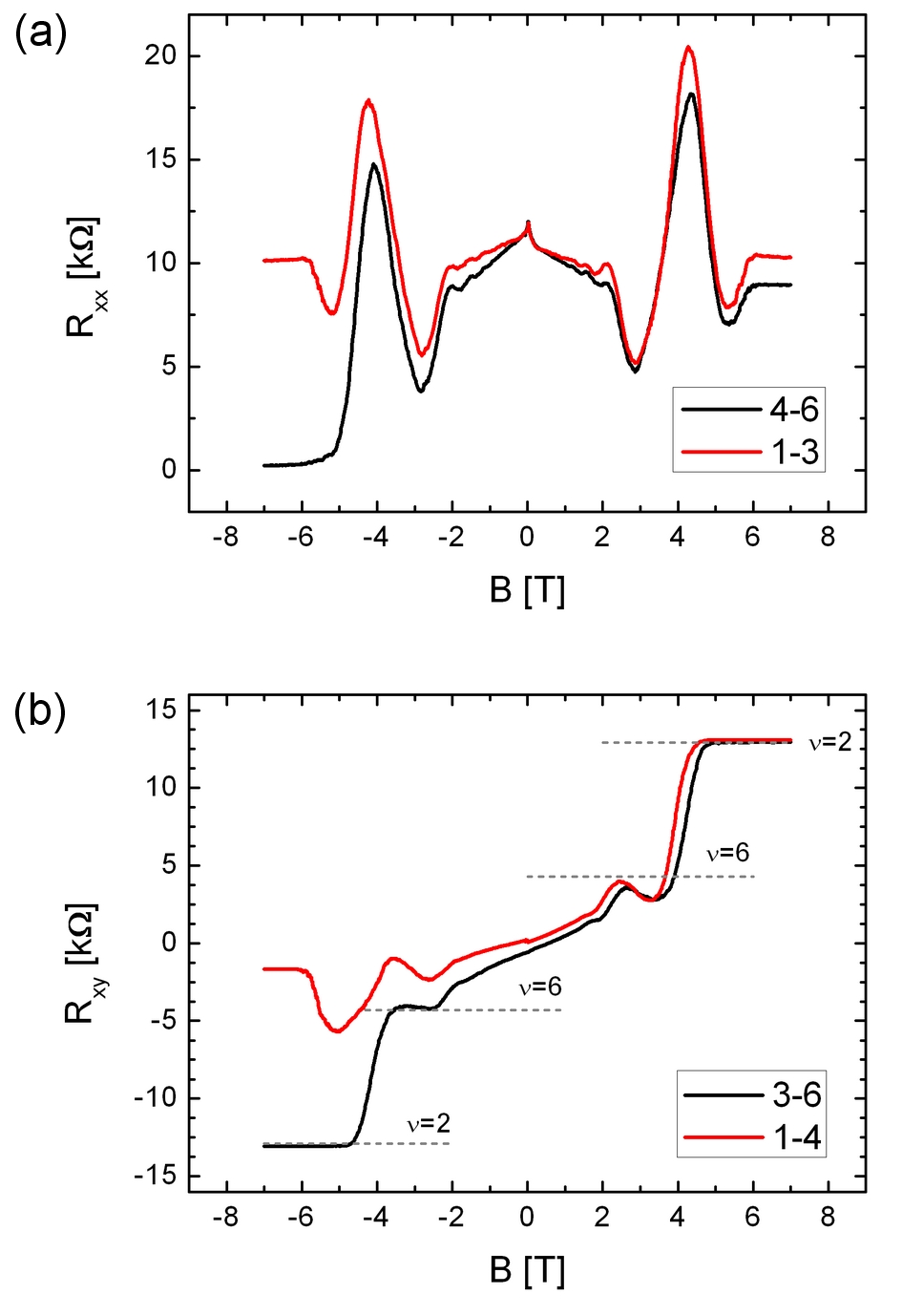} 
\caption{(Color online) Quantum Hall effect in an epitaxial graphene device grown on SiC(0001) and oriented perpendicularly to the SiC step edges: (a) Longitudinal and (b) Hall resistance as a function of magnetic field at $T=250\;mK$. The traces were measured using different contact pairs. The values of the filling factor $\nu$ are indicated.}
\label{fig3}
\end{figure}

Magnetoresistance traces of our device are shown in Fig.~\ref{fig3}. The traces of longitudinal (Fig.~\ref{fig3}(a)) and transverse (Fig.~\ref{fig3}(b)) resistance were measured using different contact pairs. For $\left| B \right| < 5~T$, the traces of the longitudinal resistance are similar, and both show the typical behavior expected for clean graphene monolayer Hall--bars, comprising a weak localization peak around zero-field, and the developing of magneto-oscillations precursory to the Shubnikov-de Haas oscillations.\cite{Iagallo2013} In the same field range, both traces of transverse resistance $R_{xy}$ show a monotonic dependence and display kinks at $\left| B\right|\approx 3\;T$. These kinks coincide with the most pronounced minima in $R_{xx}$ and correspond to a filling factor $\nu$=6.

For $\left|B\right|>5\;T$, the curves measured using different contact pairs are strongly asymmetric, and deviate significantly from the quantized values expected for homogeneous monolayer graphene Hall bars in the QH regime. Considering the longitudinal resistance, both $R_{1-3}$ and $R_{4-6}$ saturate at a value $\approx 10\;k\Omega$ for $B>0$. For $B<0$, $R_{1-3}$ saturates at the same value $\approx 10\; k\Omega$, whereas $R_{4-6}$ display a vanishing value. Correspondingly, the transverse resistance $R_{3-6}$ displays a $0.5\times h/e^{2}$-plateau for both field signs, whereas $R_{1-4}$ reaches the $0.5\times h/e^{2}$-plateau only for $B>0$, and $R_{1-4}\approx -1.6\;k\Omega$ for $B<0$.

A magnetic-field dependence similar to $R_{1-3}$ was already observed for Hall bars intersected by bilayer inclusions which are believed to shunt the transport channels at opposite sides of the bar.\cite{SchumannPRB2012,YagerNL2013,ChuaNL2014} In these structures, $R_{xx}$ was insensitive to an inversion of the direction of circulation of the channels, and it thus exhibited the same saturation values regardless of the sign of magnetic field. In contrast to these previous findings, however, here we show that $R_{4-6}$ has a remarkably asymmetric field-dependence, and is not invariant under inversion of the magnetic field. Analogously, in the transverse direction, a symmetric behavior is displayed only by $R_{3-6}$, with clear plateaus corresponding to filling factor $\nu=2$, as expected for transverse contact pairs connected by a continuous monolayer region. On the other hand, $R_{1-4}$ is markedly asymmetric, thus implying the presence of an effect which depends on the chirality of the edge-channels. We note that for $\left|B\right|>5 \; T$, all magnetoresistance curves are flat, suggesting that their values are pinned to quantized resistance plateaus.

In order to analyze our results, in the following we develop a model based on the assumption that additional channels are present in the bilayer regions that cause a shunting of the edge states circulating in the monolayer graphene. This of course requires that bilayer stripes connect continuously both sides of the Hall bar, consistently with our Raman results in Fig.~\ref{fig2}.

\begin{figure}[tbp]
\includegraphics[width=\columnwidth]{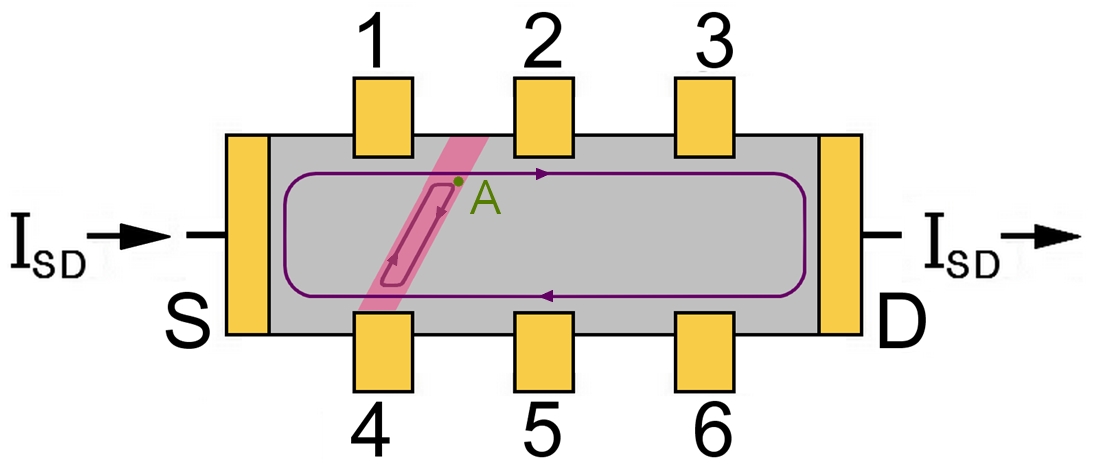} 
\caption{(Color online) Schematic describing the channel arrangement used to model the magnetoresistance. A bilayer domain (purple band) at filling factor $\nu_{B}>\nu_{M}$ connects contact 4 to the opposite side of the Hall--bar, inducing mixing of the channels propagating between $S$ and $D$ in the monolayer region.}
\label{fig4}
\end{figure}

Our quantitative analysis exploits to the B\"uttiker-Landauer formalism,\cite{ButtikerPRB1988} in an approach similar to the one applied to monolayer-bilayer planar junctions on exfoliated graphene.\cite{KiPRB2009,KiPRB2010,TianPRB2013} We consider our device as consisting of a monolayer and a bilayer region, both in a QH state, and hosting different numbers of edge states. Differently from the symmetric cases which are found in the literature,\cite{SchumannPRB2012,YagerNL2013,ChuaNL2014} the peculiar asymmetry of the curves in Fig.~\ref{fig3} imposes that the bilayer connection run between contact 4 and the other side of the device between contacts 1 and 2, as shown in Fig.~\ref{fig4}. Any other configuration fails to reproduce the observed symmetries while simultaneously remaining consistent with the Raman data. The fact that in our device the ohmic contacts stick out into the Hall bar makes this configuration more likely than for a Hall bar with recessed ohmic contacts and is probably the reason why this asymmetry was not observed before.

The arrangement considered is depicted in Fig.~\ref{fig4} for clockwise (CW) channel chirality ($B>0$) and filling factors in the monolayer ($\nu_{M}$) and in the bilayer ($\nu_{B}$) such that $\nu_{B}>\nu_{M}$. In such a case, the channels exiting contact 1 and the channels from contact 4 undergo a full equilibration while co-propagating along the upper device edge, and emerge at point $A$ at the same potential. By making use of current conservation relations, it is possible to write down the equation at each node, obtaining 
\begin{equation}
V_{4-6}=V_{1-3}=\frac{\nu_{B}-\nu_{M}}{2\nu_{B}-\nu_{M}} V_{SD}
\label{eqXXCW}
\end{equation}
for the longitudinal direction, and
\begin{equation}
V_{1-4}=V_{3-6}=\frac{\nu_{B}}{2\nu_{B}-\nu_{M}}V_{SD}
\label{eqXYCW}
\end{equation}
for the transverse direction. We note that for positive sign of the magnetic field, longitudinal voltages have identical values for opposite sides of the devices, as in the conventional quantum Hall regime. However, their value is considerably larger due to the presence of shunting channels in the bilayer.

For $B<0$, a counter-clockwise (CCW) propagation is expected and the corresponding values for the longitudinal voltage drops are
\begin{equation}
V_{4-6}=0, \;\;\;
V_{1-3}=\frac{\nu_{B}-\nu_{M}}{2\nu_{B}-\nu_{M}} V_{SD},
\label{eqXXACW}
\end{equation}
while in the transverse direction we obtain
\begin{equation}
V_{1-4}=-\frac{\nu_{M}}{2\nu_{B}-\nu_{M}} V_{SD}, \;\;\;
V_{3-6}=-\frac{\nu_{B}}{2\nu_{B}-\nu_{M}} V_{SD}.
\label{eqXYACW}
\end{equation}

\renewcommand{\arraystretch}{2}
\begin{table}[tbp]
\caption{Calculated quantum Hall resistances (in units of $h/e^{2}$) obtained from the model in Fig.~\ref{fig4}.}	
\centering				
\begin{tabular}{c c c c c}		
\hline
\hline 					
 & $R_{1-3}$ & $R_{4-6}$ & $R_{1-4}$ & $R_{3-6}$ \\ [0.5ex] 
\hline			
B$>$0 (CW) & $\frac{\nu_{B}-\nu_{M}}{\nu_{B}\nu_{M}}$ & $\frac{\nu_{B}-\nu_{M}}{\nu_{B}\nu_{M}}$ & $\frac{1}{\nu_{M}}$ & $\frac{1}{\nu_{M}}$ \\			
B$<$0 (CCW) & $\frac{\nu_{B}-\nu_{M}}{\nu_{B}\nu_{M}}$ & $0$ & $-\frac{1}{\nu_{B}}$ & $-\frac{1}{\nu_{M}}$ \\	[1ex]
\hline					
\end{tabular}
\label{table:values}			
\end{table}

The values of the resistance, calculated using the source-drain current $I_{SD}=\nu_{M}\nu_{B}/(2\nu_{B}-\nu_{M})V_{SD}$, are summarized in Table \ref{table:values} in units of $h/e^{2}$. In the following we shall focus on the case $\left|B\right|>5\;T$, in which $\nu_{M}$=2 channels propagate along the device, as inferred from the high-field values of $R_{3-6}$. By setting $\nu_{M}$=2, and using the measured value $R_{1-3}|_{-7T}=0.393\times h/e^{2}$, we obtain the filling factor in the bilayer region $\nu_{B}=9.33$. The nearest Hall plateau for bilayer graphene is at $\nu_{B}=8$.\cite{Novoselov2006} We can estimate the variation in our data by considering the difference in $R_{4-6}$ and $R_{1-3}$ at $B=+7 \; T$, which should be identical according to our model (see Table \ref{table:values}). We obtain $R_{1-3}|_{+7T}=0.397\times h/e^{2}$ and $R_{4-6}|_{+7T}=0.346\times h/e^{2}$, i.~e.~a variation in $R$ of $0.051\times h/e^{2}$. This slight difference in the traces measured with different contacts is presumably caused by an inhomogeneity of the charge density and a mixing between resistance components due to geometrical effects. Using the filling factor values $\nu_{M}$=2 and $\nu_{B}=8$, we calculate a transverse resistance value $R_{1-4}=-0.125\times h/e^{2}$, which is roughly consistent with the measured value $R_{1-4}|_{-7T}=-0.064\times h/e^{2}$ within the variation. In summary, this simple model explains the main features of our data. In particular, it fully accounts for the peculiar asymmetry of the magnetoresistance curves. Our results imply the coexistence of QH states on both monolayer and bilayer graphene grown on the same substrate.

This topic was addressed in a very recent work,\cite{ChuaNL2014} where an electrostatic model defining the domain of coexistence of QH in both monolayer graphene and bilayer inclusions was proposed. It was stated that QH conditions in mono-- and bilayer regions at large carrier density are not expected to be simultaneously present, so that the bilayer inclusions acts as dissipative shunts when the monolayer graphene is pinned at a certain filling factor. However, we argue that if monolayer and bilayer regions have different carrier densities, it is possible to have the conditions for dissipationless transport in both graphene domains. A density of $3.4 \times 10^{11}$~cm$^{-2}$ in monolayer graphene and $\nu=2$ would require a 4 times higher carrier density in bilayer graphene to reach $\nu=8$ at the same magnetic field $B$, i.~e.~a carrier density of $1.4 \times 10^{12}$~cm$^{-2}$. These values are in excellent agreement with values reported in literature for similar mono-- and bilayer samples.\cite{OPL:8246958}

\section{\label{secConc} Conclusion}

In summary, we observed an asymmetric dependence of the magnetoresistance of graphene in a Hall--bar oriented perpendicularly to the SiC(0001) step edges, which we attribute to the presence of continuous bilayer graphene stripes crossing the device. We propose a quantitative model involving the simultaneous coexistence of QH conditions in the monolayer and bilayer regions, at different filling factors. The transport channels in the bilayer inclusions are responsible for inter-channel scattering, which results in mixing of the edge-channels in the monolayer and deviations from the conventional quantum Hall effect.

\begin{acknowledgments}
The authors acknowledge financial support from the Italian Ministry of Research (MIUR-FIRB project RBID08B3FM), from the Italian Ministry of Foreign Affairs (Ministero degli Affari Esteri, Direzione Generale per la Promozione del Sistema Paese, progetto: Nanoelettronica quantistica per le tecnologie delle informazioni), and from the CNR in the framework of the agreement on scientific collaboration between CNR and JSPS (Japan), joint project title 'High--mobility graphene monolayers for novel quantum devices'. We also acknowledge funding from the European Union Seventh Framework Programme under grant agreement no.~604391 Graphene Flagship.
\end{acknowledgments}

\bibliography{MBlayer_grapheneBiblio}

\end{document}